\documentclass[journal,letterpaper]{IEEEtran}

\usepackage{cite}
\usepackage{textcomp}
\usepackage{amsmath, amssymb, amsfonts}
\usepackage{accents}
\usepackage{algorithm}
\usepackage{algorithmic}
\usepackage{color}
\usepackage{graphicx}
\usepackage{enumerate}
\usepackage{booktabs}
\usepackage{subcaption}
\usepackage{enumerate}
\usepackage{mathrsfs}
\usepackage{mathtools}
\usepackage{dsfont}
\usepackage{relsize}

\DeclareMathOperator{\tr}{tr}
\DeclareMathOperator{\minimize}{minimize}

\DeclareMathOperator{\diag}{diag}

\DeclareMathOperator{\E}{\mathsf{E}}
\DeclareMathOperator{\Cov}{\mathsf{cov}}

\DeclareMathOperator{\ProbM}{\mathsf{P}}
\DeclareMathOperator{\Prob}{\mathsf{p}}

\DeclareMathOperator{\rtx}{\mathsf{rtx}}
\DeclareMathOperator{\tx}{\mathsf{tx}}

\newtheorem{lemma}{Lemma}

\newtheorem{theorem}{Theorem}

\newtheorem{remark}{Remark}

\begin{document}
\title{Networked Control with Hybrid Automatic\\Repeat Request Protocols}

\author{Touraj Soleymani, John S. Baras, and Deniz G\"{u}nd\"{u}z \vspace{-2mm}
\thanks{Touraj Soleymani is with the City St George's School of Science and Technology, University of London, London EC1V~0HB, United Kingdom. Deniz G\"{u}nd\"{u}z is with the Department of Electrical and Electronic Engineering, Imperial College London, London SW7 2AZ, United Kingdom. John S.~Baras is with the Institute for Systems Research, University of Maryland, MD 20742, United States.}%
}

\maketitle

\begin{abstract}
We study feedback control of a dynamical process over a lossy channel equipped with a hybrid automatic repeat request protocol that connects a sensor to an actuator. The dynamical process is modeled by a Gauss--Markov process, and the lossy channel by a packet-erasure channel with ideal feedback. We suppose that data is communicated in the format of packets with negligible quantization error. In such a networked control system, whenever a packet loss occurs, there exists a tradeoff between transmitting new sensory information with a lower success probability and retransmitting previously failed sensory information with a higher success probability. In essence, an inherent tradeoff between freshness and reliability. To address this tradeoff, we consider a linear-quadratic-regulator performance index, which penalizes state deviations and control efforts over a finite horizon, and jointly design optimal policies for an encoder and a decoder, which are collocated with the sensor and the actuator, respectively. Our emphasis here lies specifically on designing switching and control policies, rather than error-correcting codes. We derive the structural properties of the optimal encoding and decoding policies. We show that the former is a threshold switching policy and the latter is a certainty-equivalent control policy. In addition, we specify the iterative equations that the encoder and the decoder need to solve in order to implement the optimal~policies.
\end{abstract}


\section{Introduction}
Networked control systems are distributed feedback systems where the underlying components, i.e., sensors, actuators, and controllers, are connected to each other via communication channels~\cite{baillieul2007X}. In these systems, \emph{wireless communication} can play an important role due to key reasons~\cite{park2017}. Notwithstanding the advantages, wireless communication presents some challenges. In particular, wireless channels, which serve to close the feedback control loops in networked control systems, are prone to noise. A direct consequence of the channel noise in real-time tasks is information loss, which can severely degrade the performance of the underlying system or can even lead to instability. It is known that reliable communication close to the capacity limit can be attained with error correction subject to infinite delay or with persistent retransmission based on feedback. In feedback control, where data is real-time, any delay more than a certain threshold is typically intolerable. Moreover, retransmission of stale information may not be favorable if fresh information can be transmitted with the same success rate instead. In this situation, the adoption of a \emph{hybrid automatic repeat request (HARQ) protocol}, which integrates error correction with retransmission, seems to be highly promising. Note that an HARQ protocol is able to effectively increase the successful detection probability of a retransmission by combining multiple copies from previously failed transmissions~\cite{giovanidis2009opt, li2017thr, lagrange2010, tripathi2003, frenger2001}. Despite the substantial research conducted on HARQ for enhancing transmission reliability in wireless communication systems, its application in networked control systems, however, has received very limited attention in the literature.

In the present article, we study feedback control of a dynamical process over a lossy channel equipped with an HARQ protocol that connects a sensor to an actuator. We suppose that data is communicated in the format of packets with negligible quantization error. Although the exact relationship between the probability of successful packet detection and the number of retransmissions in an HARQ protocol varies depending on the channel conditions and the adopted HARQ scheme, it is observed that the probability of successful packet detection generally increases with each retransmission. This suggests that, whenever a packet loss occurs in our networked control system, there exists a tradeoff between transmitting new sensory information with a lower success probability and retransmitting previously failed sensory information with a higher success probability. In essence, an inherent tradeoff between \emph{freshness} and \emph{reliability}. To address this tradeoff, we consider a linear-quadratic-regulator performance index, which penalizes state deviations and control efforts over a finite horizon, and jointly design optimal policies for an encoder and a decoder, which are collocated with the sensor and the actuator,~respectively.

\subsection{Related Work}
Earlier research have considered status updating and state estimation over communication channels with retransmissions~\cite{parag2017real, sac2018age, li2022age, yates2017timely, najm2017status, wang2020age, ceran2019average, ceran2021reinf, banerjee2023re, huang2020real, nadeem2022real, an2020harq}. In particular, status updating over a binary erasure channel was studied in \cite{parag2017real}, where the tradeoff between the protection afforded by additional redundancy and the decoding delay associated with longer codewords was analyzed, and the optimal codeword length for a few transmission schemes was obtained. This problem was further examined in~\cite{sac2018age} for two specific HARQ protocols based on finite-blocklength information theoretic bounds, where it was shown that there exists an optimal blocklength minimizing the average age and the average peak age of information. In \cite{li2022age}, status updating over a noisy channel with reactive and proactive HARQ protocols was investigated, and unified closed-form expressions for the average age and the average peak age of information were derived. In \cite{yates2017timely}, status updating over a binary erasure channel with two specific HARQ protocols was introduced, and the effect of these schemes on the transmission time of data was analyzed. In \cite{najm2017status}, status updating with the HARQ protocols used in \cite{yates2017timely} subject to random updates was studied, where it was proved that the optimal encoding policy discards the newly generated update and does not preempt the current one. In \cite{ceran2019average}, status updating over an erasure channel with an HARQ protocol subject to a frequency constraint was studied, where the structure of the optimal signal-independent encoding policy was determined and a reinforcement learning algorithm was proposed for the case when the channel statistics are unknown. This work was extended in \cite{ceran2021reinf} to a broadcast channel setting with multiple receivers. More recently, state estimation over an erasure channel with an HARQ protocol was studied in \cite{huang2020real, nadeem2022real, an2020harq}. In these works, loss functions were expressed as nonlinear functions of the age of information, and optimal signal-independent encoding policies were obtained. Note that the above studies mainly depend on the age of information, which measures the time elapsed since the generation of the last successful delivery at each time. Although the age of information is an appropriate instrument for shaping the information flow in many networked real-time systems, it has been shown in~\cite{soleymaniCUP, soleymani2024found, soleymani2024consis, kriouile2021global, maatouk2020age, erasure2023, voi2, voi, touraj-thesis, mywodespaper, soleymani2016-cdc} that for networked control systems more comprehensive instruments such as the value of information and age of incorrect information are required, and that strategies that depend purely on the age of information can lead to a degradation in the performance of networked control~systems.

\subsection{Contributions and Outline}
This study aims to contribute to the field of \emph{semantic communications}~\cite{uysal2022semantic, gunduz2022semantic}, a new communication paradigm that, for real-time networked systems, focuses on exchanging the most significant part of data by considering its contextual relevance. In particular, we would like to explore and assess the relative significance of two types of data, namely new sensory information and previously failed sensory information, sequentially in advanced communication networks, with a specific emphasis on the regulation of dynamical systems. Our main contributions are as follows. (I) We propose a novel networked control system for feedback control of a dynamical process over a lossy channel equipped with an HARQ protocol, where the objective is to minimize state deviations and control efforts over a finite horizon. The dynamical process is modeled by a Gauss--Markov process, and the lossy channel by a packet-erasure channel with ideal feedback. We formulate the problem mathematically, and derive the structural properties of the optimal encoding and decoding policies. Our emphasis here lies specifically on designing switching and control policies, rather than error-correcting codes. (II) We show that the encoding policy is a threshold switching policy that depends on the system dynamics and its realizations, and the decoding policy is a certainty-equivalent control policy with a switching filter that depends on the encoder's decision. In addition, we specify the iterative equations that the encoder and the decoder need to solve in order to implement the optimal policies. More specifically, we show that the encoder must solve the Kalman filtering equations, a mismatch linear equation, and a Bellman optimality equation, while the decoder must solve a linear filtering equation and an algebraic Riccati equation.

It is important to note that our study diverges from~\cite{parag2017real, sac2018age, li2022age, yates2017timely, najm2017status, wang2020age, ceran2019average, ceran2021reinf, banerjee2023re, huang2020real, nadeem2022real, an2020harq}, which look for an optimal signal-independent codeword length or an optimal signal-independent encoding policy by considering a performance index for status updating or state estimation, which is expressible in terms of the age of information. We here formulate a dynamic team game with two decision makers, and search for the optimal signal-dependent encoding and decoding policies simultaneously by considering a well-established performance index for feedback control, which cannot be expressed solely in terms of the age of information. Besides, our study differs from~\cite{sinopoli, schenato, gupta2007, plarre, chiuso2014, dey2017, wu2017, mo2013, huang2007, you2011, quevedo2013, elia2005, parseh2014, gupta2009d}, which investigate the impact of channel conditions, given a stability constraint or a performance index, in a setting where, at each time, the encoder transmits new sensory information with some success probability. We here examine the impact of an HARQ protocol, given a performance index, in a setting where, at each time, the encoder decides whether to transmit new sensory information with a lower success probability or retransmit previously failed sensory information with a higher success probability. 

The article is organized as follows. We formulate the problem of interest in Section~\ref{sec2}. We present our main theoretical results in Section~\ref{sec3}. Then, we present our numerical results in Section~\ref{sec5}. Finally, we conclude the article and discuss future research in Section~\ref{sec6}.

\subsection{Notation}
In the sequel, the sets of real numbers and non-negative integers are denoted by $\mathbb{R}$ and $\mathbb{N}$, respectively. For $x,y \in \mathbb{N}$ and $x \leq y$, the set $\mathbb{N}_{[x,y]}$ denotes $\{z \in \mathbb{N} | x \leq z \leq y\}$. The sequence of all vectors $x_t$, $t=p,\dots,q$, is represented by $\boldsymbol{x}_{p:q}$. For matrices $X$ and $Y$, the relations $X \succ 0$ and $Y \succeq 0$ denote that $X$ and $Y$ are positive definite and positive semi-definite, respectively. The logical~AND and logical~OR are represented by $\wedge$ and $\vee$, respectively. The indicator function of a subset $\mathcal{A}$ of a set $\mathcal{X}$ is denoted by $\mathds{1}_\mathcal{A}:\mathcal{X} \to \{0,1\}$. The product operator $\prod_{t=p}^{q} X_t$, where $X_t$ is a matrix, is defined according to the order $X_p \cdots X_q$, and is equal to one when $q<p$. The probability measure of a random variable $x$ is represented by $\mathsf{P}(x)$, its probability density or probability mass function by $\Prob(x)$, and its expected value and covariance by $\E[x]$ and $\Cov[x]$,~respectively.

\section{Problem Statement}\label{sec2}
Consider a networked control system composed of a dynamical process, a sensor with an encoder, an actuator with a decoder, and a lossy channel with an HARQ protocol that connects the sensor to the actuator. At each time~$k$, a message containing a new measurement, represented by~$\check{x}_{k}$, or a previously failed measurement, represented by $\check{x}_{k'}$ for $k' < k$, can be transmitted or retransmitted over the channel from the sensor to the actuator, where an actuation input (i.e., control input), represented by $a_{k}$, should be computed causally in real time and over a finite time horizon~$N$. We assume that the time is discretized into time slots, and the duration of each time slot~is~constant.

The lossy channel is modeled as a packet-erasure channel with ideal feedback. The decision of the encoder at time~$k$, denoted by $u_k \in \{ \tx, \rtx\}$, can be either transmitting a new measurement (i.e., $u_k = \tx$) or retransmitting a previously failed measurement (i.e., $u_k = \rtx$). The packet loss (i.e. failure) in the channel is modeled by a random variable $\gamma_{k} \in \{0,1\}$ such that $\gamma_{k} = 0$ if a packet loss occurs at time $k$, and $\gamma_{k} = 1$ otherwise. Let us introduce the variable $\tau_k$ such that if the communication at time $k-1$ failed, $\tau_k$ is equal to the time elapsed since the last $\tx$ decision was made; otherwise, $\tau_k = 0$. Note that $\tau_k$ satisfies the recursive relation
\begin{align}\label{eq:tau-dyn}
\tau_{k} = \left\{
  \begin{array}{l l}
     1, & \ \text{if} \ u_{k-1} = \tx \ \wedge \ \gamma_{k-1} = 0, \\[1.5\jot]
     \tau_{k-1} + 1, & \ \text{if} \ u_{k-1} = \rtx \ \wedge \ \gamma_{k-1} = 0, \\[1.5\jot]
     0, & \ \text{otherwise}
  \end{array} \right.	
\end{align}
for $k \in \mathbb{N}_{[0,N]}$ with initial condition $\tau_0 = 0$. The channel satisfies the input-output relation
\begin{align}\label{eq:channel}
z_{k+1} = \left\{
  \begin{array}{l l}
     \check{x}_{k}, & \ \text{if} \ u_{k} = \tx \ \wedge \ \gamma_{k} =1, \\[1.5\jot]
     \check{x}_{k-\tau_k}, & \ \text{if} \ u_{k} = \rtx \ \wedge \ \gamma_{k} =1, \\[1.5\jot]
     \mathfrak{F} , & \ \text{otherwise}
  \end{array} \right.	
\end{align}
for $k \in \mathbb{N}_{[0,N]}$ with $z_{0} = \mathfrak{F}$ by convention, where $z_{k}$ is the output of the channel and $\mathfrak{F}$ represents the occurrence of a packet loss (i.e., failure in detection). Let $\omega_k$ represent the number of communication attempts before time $k$ associated with the previously failed measurement $\check{x}_{k-\tau_k}$. It is not difficult to observe that $\omega_k$ is in fact equal to $\tau_k$. Let the packet error rate at time~$k$ associated with a measurement after $s$ retransmissions be denoted by $\lambda_k(s)$. Then, the packet error rate for communication of the new measurement $\check{x}_{k}$ at time $k$ is $\lambda_k(0)$, and that for communication of the previously failed measurement $\check{x}_{k-\tau_k}$ at time $k$ is $\lambda_k(\omega_k)$. It is assumed that the packet error rates $\lambda_k(0)$ and $\lambda_{k}(\omega_k)$ for $k \in \mathbb{N}_{[0,N]}$ are random variables (due to channel fading) forming two Markov chains; the packet error rates $\lambda_k(0)$ and $\lambda_{k}(\omega_k)$ are estimated perfectly and known at the encoder at each time $k$; the random variables $\gamma_{k}$ for $k \in \mathbb{N}_{[0,N]}$ are mutually independent given their respective packet error rates; the maximum number of communication attempts $\omega_{\max}$ for each measurement is finite; signaling effect and measurement quantization error are negligible; a transmitted or retransmitted measurement is received after one-step delay, which is fixed and independent of the packet content; and packet acknowledgments are sent back from the decoder to the encoder via an ideal feedback~channel.

The dynamical process is modeled as a partially observable Gauss--Markov process. This process satisfies the state and output equations
\begin{align}
	x_{k+1} &= A_{k} x_{k} + B_k a_k + w_{k}\label{eq:sys}\\[2\jot]
	y_{k} &= C_{k} x_{k} + v_{k} \label{eq:sens}
\end{align}
for $k \in \mathbb{N}_{[0,N]}$ with initial condition $x_{0}$, where $x_{k} \in \mathbb{R}^n$ is the state of the process, $A_{k} \in \mathbb{R}^{n \times n}$ is the state matrix, $B_k \in \mathbb{R}^{n \times m}$ is the input matrix, $a_k \in \mathbb{R}^m$ is the actuation input, $w_{k} \in \mathbb{R}^n$ is a Gaussian white noise with zero mean and covariance $W_{k} \succ 0$, $y_{k} \in \mathbb{R}^p$ is the output of the sensor, $C_{k} \in \mathbb{R}^{p \times n}$ is the output matrix, and $v_{k} \in \mathbb{R}^p$ is a Gaussian white noise with zero mean and covariance $V_{k} \succ 0$. It is assumed that the initial condition $x_{0}$ is a Gaussian vector with mean $m_{0}$ and covariance $M_{0}$, and the random variables $x_{0}$, $w_{t}$, and $v_{s}$ for $t,s \in \mathbb{N}_{[0,N]}$ are mutually independent.

The information sets of the encoder and the decoder at time~$k$ can be represented by $\mathcal{I}^{e}_k = \{ y_{t}, z_{t}, \lambda_t, a_{t-1}, u_{t-1}, \gamma_{t-1} | t \in \mathbb{N}_{[0,k]} \}$, and $\mathcal{I}^{d}_k = \{ z_{t}, \lambda_t, a_{t-1}, u_{t-1}, \gamma_{t-1} | t \in \mathbb{N}_{[0,k]} \}$ for $k \in \mathbb{N}_{[0,N]}$, respectively. At each time $k$ for $k \in \mathbb{N}_{[0,N]}$, the encoder must decide about $u_{k}$ and the decoder about $a_{k}$ based on the stochastic kernels $\ProbM(u_{k} | \mathcal{I}^{e}_k)$ and $\ProbM(a_{k} | \mathcal{I}^{d}_k)$, respectively. A coding policy profile $(\pi,\mu)$, consisting of an encoding (i.e., switching) policy~$\pi$ and a decoding (i.e., control) policy~$\mu$, is considered admissible if $\pi = \{ \ProbM(u_{k} | \mathcal{I}^{e}_k) | k \in \mathbb{N}_{[0,N]} \}$ and $\mu = \{ \ProbM(a_{k} | \mathcal{I}^{d}_k) | k \in \mathbb{N}_{[0,N]} \}$. We would like to identify the best possible solution, denoted as $(\pi^\star,\mu^\star)$, to the stochastic optimization problem
\begin{align}\label{eq:main_problem}
	\underset{\pi \in \mathcal{P}, \mu \in \mathcal{M}}{\minimize} \ \Upsilon(\pi,\mu)
\end{align}
subject to the lossy channel model in (\ref{eq:channel}), and the dynamical process model in (\ref{eq:sys}) and (\ref{eq:sens}), where $\mathcal{P}$ and $\mathcal{M}$ are the sets of admissible encoding policies and admissible decoding policies, respectively, and 
\begin{align}\label{eq:loss-function}
\Upsilon(\pi,\mu) := \frac{1}{N+1} \E \bigg [\sum_{k=0}^{N+1} x_{k}^T Q_{k} x_{k} + \sum_{k=0}^{N} a_k^T R_{k} a_k \bigg]
\end{align}
for $Q_k \succeq 0$ and $R_k \succ 0$ as weighting matrices.

\section{Main Results}\label{sec3}
The primary focus within this section is to outline our theoretical findings. Before proceeding, it is necessary to establish several definitions. We say a policy profile $(\pi^\star,\mu^\star)$ associated with the loss function $\Upsilon(\pi,\mu)$ is globally optimal (i.e., team optimal)~if
\begin{align*}
	\Upsilon(\pi^\star,\mu^\star) \leq \Upsilon(\pi,\mu), \ \text{for all } \pi \in \mathcal{P}, \mu \in \mathcal{M}.
\end{align*}
Note that globally optimal solutions express a stronger solution concept than Nash equilibria.

We define two distinct value functions $V^e_k(\mathcal{I}^e_k)$ and $V^d_k(\mathcal{I}^d_k)$ associated with the loss function $\Upsilon(\pi,\mu)$~as
\begin{align}
	V^e_k(\mathcal{I}^e_k) :=& \min_{\pi \in \mathcal{P} : \mu = \mu^\star}\E \bigg[ \sum_{t=k+1}^{N}  s_t^T \Lambda_t s_t \Big| \mathcal{I}^e_k \bigg] \label{eq:Ve-def}\\[1\jot]
	V^d_k(\mathcal{I}^d_k) :=& \min_{\mu \in \mathcal{M}: \pi = \pi^\star}\E \bigg[ \sum_{t=k}^{N} s_t^T \Lambda_t s_t \Big| \mathcal{I}^d_k \bigg] \label{eq:Vc-def}
\end{align}
for $k \in \mathbb{N}_{[0,N]}$ given a policy profile $(\pi^\star,\mu^\star)$, where $s_t =  a_t + (B_t^T S_{t+1} B_t + R_t)^{-1} B_t^T S_{t+1} A_t x_t$ and $\Lambda_t = B_t^T S_{t+1} B_t + R_t$, and $S_t \succeq 0$ obeys the algebraic Riccati equation
\begin{align}\label{eq:riccati}
S_t &= Q_t + A_t^T S_{t+1} A_t - A_t^T S_{t+1} B_t \nonumber\\[1.5\jot]
	&\qquad \qquad \times \Big(B_t^T S_{t+1} B_t + R_t \Big)^{-1} B_t^T S_{t+1} A_t
\end{align}
for $t \in \mathbb{N}_{[0,N]}$ with initial condition $S_{N+1} = Q_{N+1}$. Note that this definition takes into account the fact that the encoder's decision at time $k$ can affect the cost function only from time $k+1$ onward.

In addition, we define the innovation at the encoder $\nu_{k} := y_{k} - C_{k} \E [x_{k} | \mathcal{I}^{e}_{k-1}]$, the estimation error at the encoder based on the conditional mean $\check{e}_{k} := x_{k} - \E[x_{k} | \mathcal{I}^{e}_k]$, the estimation error at the decoder based on the conditional mean $\hat{e}_{k} := x_{k} - \E[x_{k} | \mathcal{I}^{d}_k]$, the estimation mismatch based on the conditional means $\tilde{e}_{k} := \E[x_{k} | \mathcal{I}^{e}_k] - \E[x_{k} | \mathcal{I}^{d}_k]$, and the value residual from the perspective of the encoder $\Delta_{k} := (\lambda_k(\omega_k)-\lambda_k(0)) \E[ V_{k+1}(\mathcal{I}^{e}_{k+1}) | \mathcal{I}^{e}_k, \gamma_k = 0] - (1-\lambda_k(0))\E[ V_{k+1}(\mathcal{I}^{e}_{k+1}) | \mathcal{I}^{e}_k, u_{k} = \tx, \gamma_k = 1] + (1-\lambda_k(\omega_k)) \E[ V_{k+1}(\mathcal{I}^{e}_{k+1}) | \mathcal{I}^{e}_k, u_{k} = \rtx, \gamma_k = 1]$, for $k \in \mathbb{N}_{[0,N]}$. Note that $\nu_{k}$, $\check{e}_{k}$, $\hat{e}_{k}$, $\tilde{e}_{k}$, and $\Delta_{k}$ are all computable at the encoder at each time $k$.

Our main theoretical results are presented in the next theorems, which specify the structural properties of the encoding policy $\pi^\star$ and the decoding policy $\mu^\star$ of a globally optimal solution $(\pi^\star,\mu^\star)$.

\begin{theorem}\label{thm:1}\emph{
The optimal encoding policy $\pi^\star$ in feedback control of a Gauss--Markov process over a packet-erasure channel with an HARQ protocol is determined by the threshold switching policy
\begin{align}
	u^\star_{k} = \left\{
  \begin{array}{l l}
     \tx, & \ \text{if} \ \omega_k > \omega_{\max} \vee \tau_k = 0 \vee \Omega_k \geq 0, \\[1.5\jot]
     \rtx, & \ \text{otherwise}
  \end{array} \right.
\end{align}
along with $\check{x}_{k} = \E[x_{k} | \mathcal{I}^{e}_k]$ for $k \in \mathbb{N}_{[0,N]}$, where $\Omega_k = (\lambda_k(\omega_k) - \lambda_k(0)) \tilde{e}_k^T A_k^T \Gamma_{k+1} A_k \tilde{e}_k + (1- \lambda_k(\omega_k)) \varepsilon_k^T \Gamma_{k+1} \varepsilon_k + \Delta_k$, $\Gamma_k = A_k^T S_{k+1} B_k (B_k^T S_{k+1} B_k + R_k)^{-1} B_k^T S_{k+1} A_k$, and $\varepsilon_k = \sum_{t=0}^{\tau_{k}-1} (\prod_{t'=0}^{t} A_{k-t'} ) K_{k-t} \nu_{k-t}$. This encoding policy can be expressed at each time $k$ as a function of $\tilde{e}_k$, $\boldsymbol{\nu}_{k-\tau_k+1:k}$, $\tau_k$, and $\lambda_k$. To apply this encoding policy, the following equations need to be solved online:
\begin{align}
	\check{x}_{k} &= A_{k-1} \check{x}_{k-1} + B_{k-1} a_{k-1} + K_{k} \nu_{k} \\[2.25\jot]
	P_{k} &= \Big( \Big(A_{k-1} P_{k-1} A_{k-1}^T  \nonumber \\[1\jot]
	&\qquad \qquad \quad + W_{k-1} \Big)^{-1} + C_{k}^T V_{k}^{-1} C_{k} \Big)^{-1}\\[1.25\jot]
	\tilde{e}_k &= \mathds{1}_{u_{k-1} = \tx \wedge \gamma_{k-1} = 1} K_{k} \nu_{k} \nonumber\\[0.5\jot]
	&\ +  \mathds{1}_{u_{k-1} = \rtx \wedge \gamma_{k-1} = 1} \sum_{t=0}^{\tau_{k-1}} \bigg(\prod_{t'=1}^{t} A_{k-t'} \bigg) K_{k-t} \nu_{k-t} \nonumber\\[0.25\jot]
	&\ + \mathds{1}_{\gamma_{k-1} = 0} \Big (A_{k-1} \tilde{e}_{k-1} + K_k \nu_k \Big)
\end{align}
for $k \in \mathbb{N}_{[1,N]}$ with initial conditions $\check{x}_{0} = m_{0} + K_{0} \nu_{0}$, $P_{0} = (M_{0}^{-1} + C_{0}^T V_{0}^{-1} C_{0})^{-1}$, and $\tilde{e}_{0} = K_{0} \nu_{0}$, where $K_{k} = P_{k} C_{k}^T V_{k}^{-1}$.}
\end{theorem}
\begin{IEEEproof}
See the Appendix.
\end{IEEEproof}

\begin{theorem}\label{thm:2}\emph{
The optimal decoding policy $\mu^\star$ in feedback control of a Gauss--Markov process over a packet-erasure channel with an HARQ protocol is determined by the certainty-equivalent control policy
\begin{align}
a^\star_k = -L_k \hat{x}_k
\end{align}
for $k \in \mathbb{N}_{[0,N]}$, where $L_k = (B_k^T S_{k+1} B_k + R_k)^{-1} B_k^T S_{k+1} A_k$. To apply this decoding policy, the following equation needs to be solved online:
\begin{align}
	\hat{x}_{k} &=   \mathds{1}_{u_{k-1} = \tx \wedge \gamma_{k-1} = 1} \Big( A_{k-1} \check{x}_{k-1} + B_{k-1} a_{k-1} \Big) \nonumber\\[0.25\jot]
	&\ + \mathds{1}_{u_{k-1} = \rtx \wedge \gamma_{k-1} = 1} \bigg( \bigg( \prod_{t=1}^{\tau_{k-1}+1} A_{k-t} \bigg) \check{x}_{k-\tau_{k-1}-1} \nonumber\\[0.25\jot]
	&\ + \sum_{t=0}^{\tau_{k-1}} \bigg(\prod_{t'=1}^{t} A_{k-t'} \bigg) B_{k-t-1} a_{k-t-1} \bigg) \nonumber\\[0.5\jot]
	&\ + \mathds{1}_{\gamma_{k-1} = 0} \Big( A_{k-1} \hat{x}_{k-1} + B_{k-1} a_{k-1} \Big)
\end{align}
for $k \in \mathbb{N}_{[1,N]}$ with initial condition $\hat{x}_{0} = m_{0}$, where $\hat{x}_{k} = \E[x_{k} | \mathcal{I}^{d}_k]$.}
\end{theorem}
\begin{IEEEproof}
See the Appendix.
\end{IEEEproof}

\begin{remark}
The structural results in Theorems~\ref{thm:1}~and~\ref{thm:2} certify the existence of a globally optimal solution composed of a threshold switching policy that depends on the system dynamics and its realizations, and a certainty-equivalent control policy with a switching filter that depends on the encoder's decision. Note that these policies can be designed completely separately. Moreover, the results assert that the optimal encoding policy transmits the encoder's current MMSE state estimate, i.e., $\check{x}_k$, only if $\omega_k > \omega_{\max} \vee \tau_k = 0 \vee \Omega_{k} \geq 0$, and retransmits the encoder's previously failed MMSE state estimate, i.e., $\check{x}_{k-\tau_k}$, otherwise; and that the optimal decoding policy is constructed by incorporating the decoder's current MMSE state estimate, i.e., $\hat{x}_{k}$, into the corresponding optimal state-feedback control~policy.
\end{remark}

\begin{remark}
Note that the computational complexity of the solution $(\pi^\star, \mu^\star)$ is the same as that of solving (\ref{eq:Ve-def}). As reflected in the proof of Theorem~1, $V_{k}(\mathcal{I}^e_k)$ is a function of $\tilde{e}_k$, $\boldsymbol{\nu}_{k-\tau_k+1:k}$, $\tau_k$, and $\lambda_k$. Now, if $\tau_k$ is bounded by $\tau_{\max}$, and $\tilde{e}_k$, $\nu_t$ for $t \in \mathbb{N}_{[k-\tau_k+1,k]}$, and $\lambda_k$ are discretized in grids with $c_1^{n}$, $c_2^{p}$, and $c_3$ points, respectively, and the expected value is expressed based on a weighted sum of $c_4$ samples, the computational complexity is then $\mathcal{O}(N c_1^n c_2^{p \times (\tau_{\max}-1)} c_3 c_4 \tau_{\max})$. This can be expensive especially when $n$ and $p$ are large. In practice, $\Omega_{k}$ can be approximated based on the one-step lookahead algorithm (see, e.g., \cite{bertsekas1995DP}). Applying this procedure, we find $\Omega_{k} \simeq (\lambda_k(\omega_k) - \lambda_k(0)) \tilde{e}_k^T A_k^T \Gamma_{k+1} A_k \tilde{e}_k + (1- \lambda_k(\omega_k)) \varepsilon_k^T \Gamma_{k+1} \varepsilon_k$, which is quadratic in terms of $\tilde{e}_k$~and~$\varepsilon_k$. Following the definitions of $\tilde{e}_k$ and $\varepsilon_k$, we observe that these variables are affected by the age of information at the decoder, but they cannot be expressed solely in terms of this metric, as they in general depend on the system dynamics and its realizations.
\end{remark}

\begin{figure}[t!]
\centering
  \includegraphics[width=.99\linewidth]{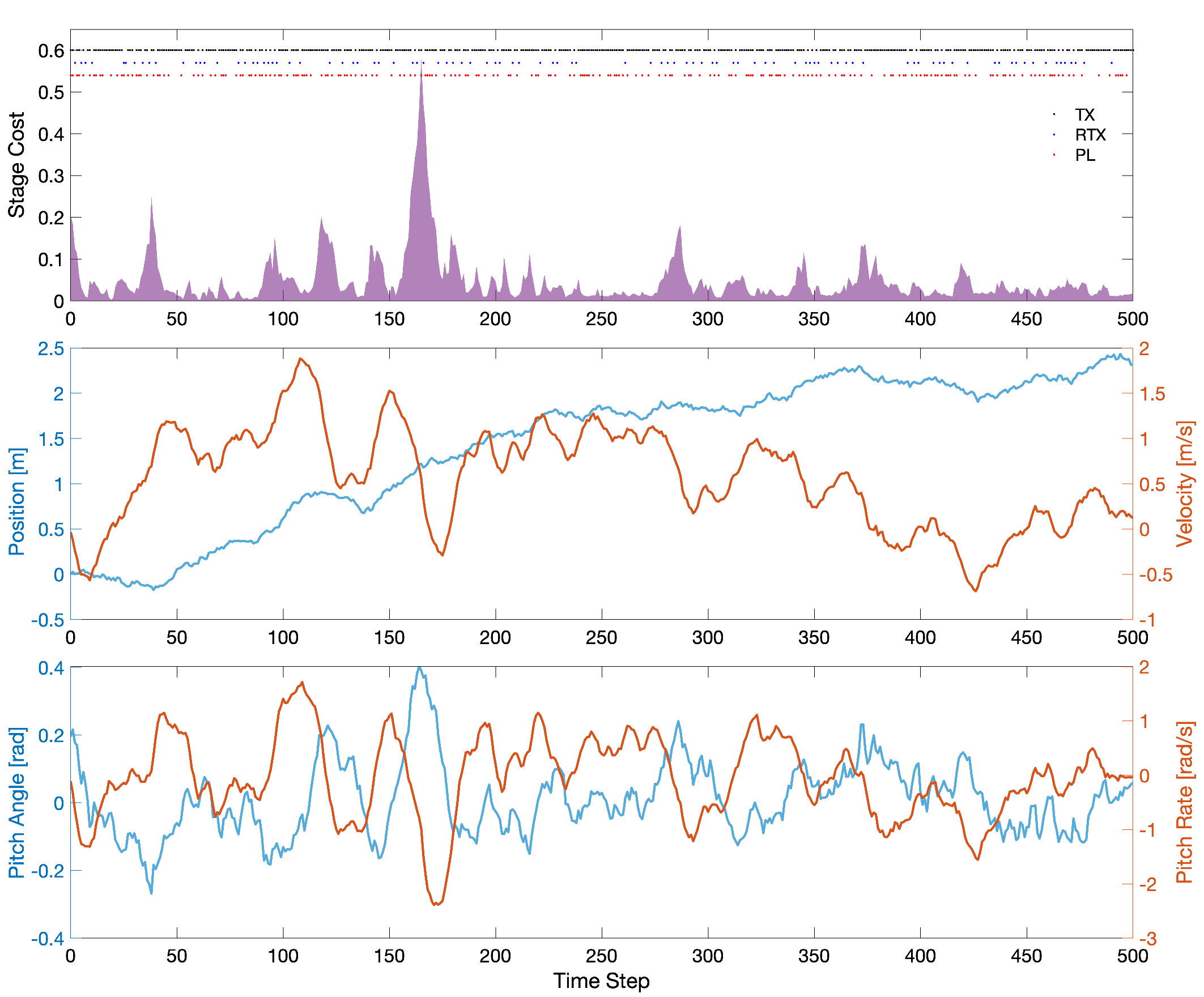}
  \caption{Simulation results for a networked control system with an HARQ protocol.}
  \label{fig:trajectories}
\end{figure}

\section{Numerical Results}\label{sec5}
In this section, we present a numerical example that can demonstrate our theoretical findings. Consider an inverted pendulum on a cart. For this dynamical process, the continuous-time equations of motion linearized around the unstable equilibrium are given~by
\begin{align*}
	(M+m) \ddot{x} + b \dot{x} - m l \ddot{\phi} = a,\\[2.25\jot]
	(I + m l^2) \ddot{\phi} - m g l \phi = m l \ddot{x}
\end{align*}
where $x$ is the position of the cart, $\phi$ is the pitch angle of the pendulum, $a$ is the force applied to the cart, $M = 0.5 \ \text{kg}$ is the mass of the cart, $m = 0.2 \ \text{kg}$ is the mass of the pendulum, $b = 0.1 \ \text{N/m/sec}$ is the coefficient of friction for the cart, $l = 0.3 \ \text{m}$ is the distance from the pivot to the pendulum's center of mass, $I = 0.006 \ \text{kg.m$^2$}$ is the moment of inertia of the pendulum, and $g = 9.81 \ \text{m/s$^2$}$ is the gravity. Suppose that an external sensor measures the position and the pitch angle at each time. In our example, the state and output equations of the dynamical process of the form~(\ref{eq:sys}) and (\ref{eq:sens}), the input-output relation of the channel of the form (\ref{eq:channel}), and the loss function of the form (\ref{eq:main_problem}) are specified by the state matrix $A_k = 10^{-4} \times [10000, 100, 1, 0 ; 0, 9982, 267, 1 ; 0, 0, 10016, 100 ; 0, -45, 3122,$ $10016]$, the input matrix $B_k = [0.0001; 0.0182; 0.0002; 0.0454]$, the output matrix $C_k = [1, 0, 0, 0; 0, 0, 1, 0]$, the process noise covariance $W_k = 10^{-4} \times [6, 3, 1, 6 ; 3, 8, 3, 4 ; 1, 3, 7, 6; 6, 4, 6, 31]$, the sensor noise covariance $V_k = 10^{-4} \times [20, 0; 0, 10]$ for $k \in \mathbb{N}_{[0,N]}$, the mean and the covariance of the initial condition $m_0 = [0; 0; 0.2; 0]$ and $M_0 = 10W_k$, the maximum number of retransmissions $\omega_{\max} = 1$, the packet error rates $\lambda_k(0) = 0.5$ and $\lambda_k(1) = 0.05$ for $k \in \mathbb{N}_{[0,N]}$, the weighting matrices $Q_k = \diag\{1,1,1000,1\}$ for $k \in \mathbb{N}_{[0,N+1]}$ and $R_k = 1$ for $k \in \mathbb{N}_{[0,N]}$, and the time horizon $N = 500$.

The simulation results corresponding to a realization of this networked control system based on the optimal coding policy profile with an HARQ protocol are illustrated in Fig.~1. In particular, in the top diagram, the solid curve represents the stage cost trajectory\footnote{Note that the stage cost at each time $k$ is defined as $(x_k^T Q_k x_k + a_k^T R_k a_k)/(N+1)$.}, the black dots represent the transmission (TX) time instants, the blue dots represent the retransmission (RTX) time instants, and the red dots represent the packet loss (PL) time instants. In this experiment, the total number of transmissions was $404$, the total number of retransmissions was $97$, and the total number of packet losses was $210$. In the middle diagram, the blue curve represents the cart's position trajectory and the red curve represents the cart's velocity trajectory. We can observe that the cart's position and velocity remained satisfactorily bounded within the ranges $[-0.17 \ \text{m}, 2.43 \ \text{m}]$ and $[-0.69 \ \text{m/s}, 1.88 \ \text{m/s}]$, respectively. Finally, in the bottom diagram, the blue curve represents the pendulum's pitch angle trajectory and the red curve represents the pendulum's pitch rate trajectory. We can observe that the pendulum's pitch angle and pitch rate remained satisfactorily bounded within the ranges $[-0.27 \ \text{rad}, 0.40 \ \text{rad}]$ and $[-2.39 \ \text{rad/s}, 1.71 \ \text{rad/s}]$,~respectively. This outcome reaffirms the effectiveness of the networked control system design proposed in our study.

\section{Conclusions}\label{sec6}
This article focused on the examination of feedback control of a Gauss--Markov process over a packet-erasure channel with an HARQ protocol and ideal feedback. We observed that, in this networked control system, whenever a packet loss occurs, there exists an inherent tradeoff between transmitting the encoder's current MMSE state estimate with a lower success probability and retransmitting the encoder's previously failed MMSE state estimate with a higher success probability. Our objective was to obtain optimal encoding and decoding policies that minimize a loss function penalizing state deviations and control efforts over a finite horizon. We derived the structural properties of the optimal policies, and specified the iterative equations that the encoder and the decoder need to solve in order to implement these policies. We suggest that future research should consider the adoption of reinforcement learning algorithms for networked control systems equipped with HARQ protocols in scenarios where the parameters and the statistics of the dynamical process and the communication channel are~unknown.

\section*{The Appendix}
This section is dedicated to the derivation of the main results. In the next two lemmas, we first characterize the recursive equations that the optimal estimators at the encoder and the decoder must~satisfy.

\begin{lemma}\label{lem:est-encoder}\emph{
The optimal estimator minimizing the MMSE at the encoder satisfies the recursive equations
\begin{align}
	\check{x}_{k} &= m_{k} + K_{k} \Big( y_{k} - C_{k} m_{k} \Big) \label{eq:est-KF-xhat} \\[1.75\jot]
	m_{k} &= A_{k-1} \check{x}_{k-1} + B_{k-1} a_{k-1} \label{eq:est-KF-m} \\[1.75\jot]
	P_{k} &= \Big( M_{k}^{-1} + C_{k}^T V_{k}^{-1} C_{k} \Big)^{-1}\\[1.75\jot]
	M_{k} &= A_{k-1} P_{k-1} A_{k-1}^T + W_{k-1}
\end{align}
for $k \in \mathbb{N}_{[1,N]}$ with initial conditions $\check{x}_0 = m_0 + K_0(y_0 - C_0 m_0)$ and $P_0 = (M_0^{-1} + C_{0}^T V_{0}^{-1} C_{0})^{-1}$, where $m_k = \E[ x_k | \mathcal{I}^e_{k-1}]$, $P_k = \Cov[x_k | \mathcal{I}^e_{k}]$, $M_k = \Cov[x_k | \mathcal{I}^e_{k-1}]$, and $K_{k} = P_{k} C_{k}^T V_{k}^{-1}$.}
\end{lemma}

\begin{IEEEproof}
Given the information set $\mathcal{I}^e_k$, the optimal estimator at the encoder must satisfy the Kalman filter equations for the conditional mean and the conditional covariance. For detailed derivation of these equations see e.g., \cite{stengel1994}.
\end{IEEEproof}

\begin{lemma}\label{lem:est-decoder}\emph{
The optimal estimator minimizing the MMSE at the decoder satisfies the recursive equation
\begin{align}\label{est-monitor}
	&\hat{x}_{k} =   \mathds{1}_{u_{k-1} = \tx \wedge \gamma_{k-1} = 1} \Big( A_{k-1} \check{x}_{k-1} + B_{k-1} a_{k-1} \Big) \nonumber\\[0\jot]
	&\qquad + \mathds{1}_{u_{k-1} = \rtx \wedge \gamma_{k-1} = 1} \bigg( \bigg( \prod_{t=1}^{\tau_{k-1}+1} A_{k-t} \bigg) \check{x}_{k-\tau_{k-1}-1} \nonumber\\[0\jot]
	&\qquad + \sum_{t=0}^{\tau_{k-1}} \bigg(\prod_{t'=1}^{t} A_{k-t'} \bigg) B_{k-t-1} a_{k-t-1} \bigg) \nonumber\\[1\jot]
	&\qquad + \mathds{1}_{\gamma_{k-1} = 0} \Big( A_{k-1} \hat{x}_{k-1} + B_{k-1} a_{k-1} \Big)
\end{align}
for $k \in \mathbb{N}_{[1,N]}$ with initial condition $\hat{x}_0 = m_0$.
}
\end{lemma}

\begin{IEEEproof}
Writing $x_k$ in terms of $x_{k-1}$ based on (3) in our paper, and taking expectation given the information set $\mathcal{I}^d_k$, we obtain
\begin{align}\label{eq:propagation2}
\E \Big[ x_k \big| \mathcal{I}^d_k \Big] = A_{k-1} \E \Big[x_{k-1} \big| \mathcal{I}_k^d \Big] + B_{k-1} a_{k-1}
\end{align}
for $k \in \mathbb{N}_{[1,N]}$ as $\E[w_{k-1}|\mathcal{I}^d_k] = 0$. If $\gamma_{k-1} = 0$, regardless of $u_{k-1}$, we have $z_k = \mathfrak{F}$. Note that $\E[x_{k-1} | \mathcal{I}^d_{k}] =\E[x_{k-1} | \mathcal{I}^d_{k-1}, z_k = \mathfrak{F}, \lambda_k, a_{k-1}, u_{k-1}, \gamma_{k-1} = 0] = \E[x_{k-1} | \mathcal{I}^d_{k-1}] = \hat{x}_{k-1}$. Therefore, using (\ref{eq:propagation2}), when $\gamma_{k-1} = 0$, regardless of $u_{k-1}$, we get
\begin{align}\label{eq:erasure-update1}
	\E \Big[x_k \big| \mathcal{I}^d_k \Big] =  A_{k-1} \hat{x}_{k-1} + B_{k-1} a_{k-1}
\end{align}
for $k \in \mathbb{N}_{[1,N]}$.

However, if $u_{k-1} = \tx$ and $\gamma_{k-1} = 1$, we have $z_k = \check{x}_{k-1}$. In this case, we get $\E[x_{k-1} | \mathcal{I}^d_k] = \E[x_{k-1} | \mathcal{I}^d_{k-1}, z_k = \check{x}_{k-1}, \lambda_k, a_{k-1}, u_{k-1} = \tx, \gamma_{k-1} = 1] = \E[x_{k-1} | \check{x}_{k-1}, P_{k-1}] = \check{x}_{k-1}$ as $\{ \check{x}_{k-1}, P_{k-1} \}$ is a sufficient statistic of $\mathcal{I}^d_k$ with respect to $x_{k-1}$. Hence, using (\ref{eq:propagation2}), when $u_{k-1} = \tx$ and $\gamma_{k-1} = 1$, we~get
\begin{align}\label{eq:erasure-update2}
	\E \Big[x_k \big| \mathcal{I}^d_k \Big] =  A_{k-1} \check{x}_{k-1} + B_{k-1} a_{k-1}
\end{align}
for $k \in \mathbb{N}_{[1,N]}$.

Furthermore, writing $x_k$ in terms of $x_{k-\tau_{k-1}-1}$ based on~(3) in our paper, and taking expectation given the information set $\mathcal{I}^d_k$, we obtain
\begin{align}\label{eq:propagation}
\E \Big[ x_k \big| \mathcal{I}^d_k \Big] &= \bigg( \prod_{t=1}^{\tau_{k-1}+1} A_{k-t} \bigg) \E \Big[x_{k-\tau_{k-1}-1} \big| \mathcal{I}_k^d \Big] \nonumber\\[0.5\jot]
	&\quad + \sum_{t=0}^{\tau_{k-1}} \bigg(\prod_{t'=1}^{t} A_{k-t'} \bigg) B_{k-t-1} a_{k-t-1}
\end{align}
for $k \in \mathbb{N}_{[1,N]}$ as $ \sum_{t=0}^{\tau_{k-1}} (\prod_{t'=1}^{t} A_{k-t'}) \E[ w_{k-t-1} | \mathcal{I}^d_k] = 0$. If $u_{k-1} = \rtx$ and $\gamma_{k-1} = 1$, we have $z_k = \check{x}_{k-\tau_{k-1}-1}$. In this case, we get $\E[x_{k-\tau_{k-1}-1} | \mathcal{I}^d_k] = \E[x_{k-\tau_{k-1}-1} | \mathcal{I}^d_{k-1}, z_k = \check{x}_{k-\tau_{k-1}-1}, \lambda_k, a_{k-1}, u_{k-1} = \rtx, \gamma_{k-1} = 1] = \E[x_{k-\tau_{k-1}-1} | \check{x}_{k-\tau_{k-1}-1}, P_{k-\tau_{k-1}-1}] = \check{x}_{k-\tau_{k-1}-1}$ as $\{ \check{x}_{k-\tau_{k-1}-1}, P_{k-\tau_{k-1}-1} \}$ is a sufficient statistic of $\mathcal{I}^d_k$ with respect to $x_{k-\tau_{k-1}-1}$. Hence, using (\ref{eq:propagation}), when $u_{k-1} = \rtx$ and $\gamma_{k-1} = 1$, we get
\begin{align}\label{eq:update0}
	\E \Big[x_k \big| \mathcal{I}^d_k \Big] &= \bigg( \prod_{t=1}^{\tau_{k-1}+1} A_{k-t} \bigg) \check{x}_{k-\tau_{k-1}-1} \nonumber\\[0.5\jot]
	&\quad + \sum_{t=0}^{\tau_{k-1}} \bigg(\prod_{t'=1}^{t} A_{k-t'} \bigg) B_{k-t-1} a_{k-t-1}
\end{align}
for $k \in \mathbb{N}_{[1,N]}$.

We obtain (\ref{est-monitor}) by combining (\ref{eq:erasure-update1}), (\ref{eq:erasure-update2}), and (\ref{eq:update0}). Note that the initial condition is $\E[x_{0}] = m_{0}$ because no measurement is available at the decoder at time $k = 0$.
\end{IEEEproof}

In addition, in the next two lemmas, we provide some properties of the estimation error at the decoder and the estimation mismatch.

\begin{lemma}\label{lem:error-dyn}\emph{
The estimation error at the decoder satisfies the recursive equation
\begin{align}\label{eq:error-dyn}
	\hat{e}_k &= \mathds{1}_{u_{k-1} = \tx \wedge \gamma_{k-1} = 1} \Big (A_{k-1} \check{e}_{k-1} + w_{k-1} \Big) \nonumber\\[0.5\jot]
	& +  \mathds{1}_{u_{k-1} = \rtx \wedge \gamma_{k-1} = 1} \sum_{t=0}^{\tau_{k-1}} \bigg(\prod_{t'=1}^{t} A_{k-t'} \bigg) w_{k-t-1} \nonumber\\[0.5\jot]
	& + \mathds{1}_{\gamma_{k-1} = 0} \Big (A_{k-1} \hat{e}_{k-1} + w_{k-1} \Big)
\end{align}
for $k \in \mathbb{N}_{[1,N]}$ with initial condition $\hat{e}_0 = \hat{x}_0 - m_0$.}
\end{lemma}

\begin{IEEEproof}
Using (3) in our paper, we can write $x_k$ in terms of $x_{k-1}$~as
\begin{align}\label{eq:filter-1-x}
	x_k &= A_{k-1} x_{k-1} + B_{k-1} a_{k-1} + w_{k-1}
\end{align}
for $k \in \mathbb{N}_{[1,N]}$ with initial condition $x_0$, and in terms of $x_{k-\tau_{k-1}-1}$~as
\begin{align}\label{eq:filter-2-x}
	x_k &= \bigg(\prod_{t=1}^{\tau_{k-1}+1} A_{k-t}\bigg) x_{k-\tau_{k-1}-1} \nonumber\\
	&\quad + \sum_{t=0}^{\tau_{k-1}} \bigg(\prod_{t'=1}^{t} A_{k-t'} \bigg) B_{k-t-1} a_{k-t-1} \nonumber\\
	&\quad + \sum_{t=0}^{\tau_{k-1}} \bigg(\prod_{t'=1}^{t} A_{k-t'}\bigg) w_{k-t-1}
\end{align}
for $k \in \mathbb{N}_{[1,N]}$ with initial condition $x_{0}$.

Moreover, from the law of total probability, $\Prob(u_{k-1} = \tx \ \wedge \ \gamma_{k-1} = 1) + \Prob(u_{k-1} = \rtx \ \wedge \ \gamma_{k-1} = 1) + \Prob(\gamma_{k-1} = 0) = 1$. Thus, using (\ref{eq:filter-1-x}) and (\ref{eq:filter-2-x}), we can~get
\begin{align}\label{eq:filter-3-x}
	x_k &= \mathds{1}_{u_{k-1} = \tx \wedge \gamma_{k-1} = 1}  \Big( A_{k-1} x_{k-1} + B_{k-1} a_{k-1} + w_{k-1} \Big) \nonumber\\[1\jot]
	&\quad + \mathds{1}_{u_{k-1} = \rtx \wedge \gamma_{k-1} = 1} \bigg(\bigg(\prod_{t=1}^{\tau_{k-1}+1} A_{k-t}\bigg) x_{k-\tau_{k-1}-1} \nonumber\\
	&\quad + \sum_{t=0}^{\tau_{k-1}} \bigg(\prod_{t'=1}^{t} A_{k-t'} \bigg) B_{k-t-1} a_{k-t-1} \nonumber\\[0.5\jot]
	&\quad + \sum_{t=0}^{\tau_{k-1}} \bigg(\prod_{t'=1}^{t} A_{k-t'}\bigg) w_{k-t-1} \bigg) \nonumber\\[2\jot]
	&\quad + \mathds{1}_{\gamma_{k-1} = 0} \Big( A_{k-1} x_{k-1} + B_{k-1} a_{k-1} + w_{k-1} \Big)
\end{align}
for $k \in \mathbb{N}_{[1,N]}$ with initial condition $x_{0}$.

By definition, $\hat{e}_k = x_k - \hat{x}_k$. Therefore, we obtain (\ref{eq:error-dyn}) by subtracting (\ref{est-monitor}) from (\ref{eq:filter-3-x}). The initial condition $\hat{e}_{0}$ is also obtained by subtracting $\hat{x}_{0} = m_0$ from $x_{0}$.
\end{IEEEproof}

\begin{lemma}\label{lem:mismatch-dyn}\emph{
The estimation mismatch satisfies the recursive equation
\begin{align}\label{eq:mismatch-dyn}
	\tilde{e}_k &= \mathds{1}_{u_{k-1} = \tx \wedge \gamma_{k-1} = 1} K_{k} \nu_{k} \nonumber\\[1.75\jot]
	& +  \mathds{1}_{u_{k-1} = \rtx \wedge \gamma_{k-1} = 1} \sum_{t=0}^{\tau_{k-1}} \bigg(\prod_{t'=1}^{t} A_{k-t'} \bigg) K_{k-t} \nu_{k-t} \nonumber\\[0.5\jot]
	& + \mathds{1}_{\gamma_{k-1} = 0} \Big (A_{k-1} \tilde{e}_{k-1} + K_k \nu_k \Big)
\end{align}
for $k \in \mathbb{N}_{[1,N]}$ with initial condition $\tilde{e}_0 = K_0 \nu_0$.}
\end{lemma}

\begin{IEEEproof}
From the definition of the innovation $\nu_k$ and the state estimate $\check{x}_k$, we find that
\begin{align}\label{eq:def-inno}
	\nu_k &= y_k - C_k \Big(A_{k-1} \check{x}_{k-1} + B_{k-1} a_{k-1}\Big)
\end{align}
with the exception $\nu_0 = y_0 - C_0 m_0$. Note that $\nu_{k}$ is a white Gaussian noise with zero mean and covariance $N_{k} = C_{k} M_k C_{k}^T + V_{k}$.

Using (\ref{eq:est-KF-xhat}), (\ref{eq:est-KF-m}), and (\ref{eq:def-inno}), we can write $\check{x}_k$ in terms of $\check{x}_{k-1}$~as
\begin{align}\label{eq:filter-1-mismatch}
	\check{x}_k &= A_{k-1} \check{x}_{k-1} + B_{k-1} a_{k-1} + K_k \nu_k
\end{align}
for $k \in \mathbb{N}_{[1,N]}$ with initial condition $\check{x}_0 = m_0 + K_0 \nu_0$, and in terms of $\check{x}_{k-\tau_{k-1}-1}$~as
\begin{align}\label{eq:filter-2-mismatch}
	\check{x}_k &= \bigg(\prod_{t=1}^{\tau_{k-1}+1} A_{k-t}\bigg) \check{x}_{k-\tau_{k-1}-1} \nonumber\\
	&\quad + \sum_{t=0}^{\tau_{k-1}} \bigg(\prod_{t'=1}^{t} A_{k-t'} \bigg) B_{k-t-1} a_{k-t-1} \nonumber\\
	&\quad + \sum_{t=0}^{\tau_{k-1}} \bigg(\prod_{t'=1}^{t} A_{k-t'}\bigg) K_{k-t} \nu_{k-t}
\end{align}
for $k \in \mathbb{N}_{[1,N]}$ with initial condition $\check{x}_{0} = m_0 + K_0 \nu_0$.

Moreover, from the law of total probability, we have $\Prob(u_{k-1} = \tx \ \wedge \ \gamma_{k-1} = 1) + \Prob(u_{k-1} = \rtx \ \wedge \ \gamma_{k-1} = 1) + \Prob(\gamma_{k-1} = 0) = 1$. Thus, using (\ref{eq:filter-1-mismatch}) and (\ref{eq:filter-2-mismatch}), we can~get
\begin{align}\label{eq:filter-3-mismatch}
	\check{x}_k &= \mathds{1}_{u_{k-1} = \tx \wedge \gamma_{k-1} = 1}  \Big( A_{k-1} \check{x}_{k-1} + B_{k-1} a_{k-1} + K_k \nu_k \Big) \nonumber\\[1\jot]
	&\quad + \mathds{1}_{u_{k-1} = \rtx \wedge \gamma_{k-1} = 1} \bigg(\bigg(\prod_{t=1}^{\tau_{k-1}+1} A_{k-t}\bigg) \check{x}_{k-\tau_{k-1}-1} \nonumber\\
	&\quad + \sum_{t=0}^{\tau_{k-1}} \bigg(\prod_{t'=1}^{t} A_{k-t'} \bigg) B_{k-t-1} a_{k-t-1} \nonumber\\[0.5\jot]
	&\quad + \sum_{t=0}^{\tau_{k-1}} \bigg(\prod_{t'=0}^{t} A_{k-t'}\bigg) K_{k-t} \nu_{k-t} \bigg) \nonumber\\[1.75\jot]
	&\quad + \mathds{1}_{\gamma_{k-1} = 0} \Big( A_{k-1} \check{x}_{k-1} + B_{k-1} a_{k-1} + K_k \nu_k \Big)
\end{align}
for $k \in \mathbb{N}_{[1,N]}$ with initial condition $\check{x}_{0} = m_0$.

By definition, $\tilde{e}_k = \check{x}_k - \hat{x}_k$. Therefore, we obtain (\ref{eq:mismatch-dyn}) by subtracting (\ref{est-monitor}) from (\ref{eq:filter-3-mismatch}). The initial condition $\tilde{e}_{0}$ is also obtained by subtracting $\hat{x}_{0} = m_0$ from $\check{x}_{0} = m_0 + K_0 \nu_0$. 
\end{IEEEproof}

We now present the proofs of Theorems~1 and 2.

\begin{IEEEproof}
Given any optimal switching policy implemented, the optimal value that minimizes the mean square error at the decoder at time~$k$ is the conditional mean $\E[x_{k} | \mathcal{I}^{d}_k]$. Therefore, without loss of optimality, the decoder at each time~$k$ can use $\hat{x}_{k} = \E[x_{k} | \mathcal{I}^{d}_k]$ as the state estimate. Besides, the conditional mean $\E[x_{k} | \mathcal{I}^{e}_k]$ combines all current and previous outputs of the sensor that are accessible to the encoder at time $k$. This implies that if this new message is transmitted by the encoder at time $k$, from the MMSE perspective, the decoder is able to develop a state estimate upon the successful receipt of the message at time $k+1$ that would be the same if it had all the previous outputs of the sensor until time $k$, which is the best possible case for the decoder. Moreover, given an HARQ protocol, when the last transmitted message fails, aside from transmitting a new message, the encoder at time~$k$ has an additional choice to retransmit a previously failed message. This message, following the definition of $\tau_k$, is in fact $\E[x_{k-\tau_k} | \mathcal{I}^{e}_{k-\tau_k}]$. Therefore, without loss of optimality, the encoder at each time $k$ can choose to transmit either $\check{x}_{k} = \E[x_{k} | \mathcal{I}^{e}_k]$ or $\check{x}_{k-\tau_k} = \E[x_{k-\tau_k} | \mathcal{I}^{e}_{k-\tau_k}]$.

Let $(\pi^o,\mu^o)$ denote a policy profile in the set of globally optimal solutions. It is evident that this set cannot be empty. We prove that the policy profile $(\pi^\star,\mu^\star)$ in the claim is globally optimal by showing that $\Upsilon(\pi^\star,\mu^\star)$ cannot be greater than $\Upsilon(\pi^o,\mu^o)$. Our proof is structured in the following way:
\begin{align}
	\Upsilon(\pi^o,\mu^o) = \Upsilon(\pi^n,\mu^o) \geq \Upsilon(\pi^n,\mu^c) \geq \Upsilon(\pi^\star,\mu^\star) .
\end{align}
In particular, we first find an innovation-based switching policy $\pi^n$ such that $\Upsilon(\pi^n,\mu^o) = \Upsilon(\pi^o,\mu^o)$. Then, we derive a certainty-equivalent control policy $\mu^c$ such that $\Upsilon(\pi^n,\mu^c) \leq \Upsilon(\pi^n,\mu^o)$. Finally, we show that for the policy profile in the claim we have $\Upsilon(\pi^\star,\mu^\star) \leq \Upsilon(\pi^n,\mu^c)$. Throughout our analysis, without loss of generality, we assume that $m_0 = 0$. Similar arguments can be made for $m_0 \neq 0$ following a coordinate transformation.

In the first step of the proof, we will show that, given the control policy $\mu^o$, we can find an innovation-based switching policy $\pi^n$ that is equivalent to the switching policy $\pi^o$. Note that a switching policy in the context of our problem is innovation-based if it depends on $\boldsymbol{\nu}_{0:k}$ instead of $\boldsymbol{y}_{0:k}$ and $\boldsymbol{z}_{0:k}$ at each time $k$. From the definition of $\nu_k$, we get
\begin{align}
	\boldsymbol{y}_k = \boldsymbol{\nu}_k + E_k \check{\boldsymbol{x}}_{k-1} + F_k \boldsymbol{a}_{k-1}
\end{align}
where $E_k$ and $F_k$ are matrices of proper dimensions. From (\ref{eq:est-KF-xhat}) and (\ref{eq:est-KF-m}), we find that
\begin{align}
	\check{\boldsymbol{x}}_k = G_k \boldsymbol{\nu}_k + H_k \boldsymbol{a}_{k-1}
\end{align}
where $G_k$ and $H_k$ are matrices of proper dimensions. Furthermore, from (2) in our paper, we know that $\boldsymbol{z}_k$ is a function of $\check{\boldsymbol{x}}_{k-1}$, $\boldsymbol{u}_{k-1}$, and $\boldsymbol{\gamma}_{k-1}$. As a result, it is possible to write
\begin{align*}
\Prob_{\pi^o}(u_k | \mathcal{I}^e_k) &= \Prob_{\pi^o} \Big(u_k \big| \boldsymbol{\nu}_k, \boldsymbol{\lambda}_{k}, \boldsymbol{u}_{k-1}, \boldsymbol{a}_{k-1}, \boldsymbol{\gamma}_{k-1} \Big) \\[1.5\jot]
\Prob_{\mu^o}(a_k | \mathcal{I}^d_k) &= \Prob_{\mu^o} \Big(a_k \big| \boldsymbol{\nu}_{k-1}, \boldsymbol{\lambda}_{k}, \boldsymbol{u}_{k-1}, \boldsymbol{a}_{k-1}, \boldsymbol{\gamma}_{k-1} \Big).
\end{align*}
Accordingly, any realizations of $u_k$ and $a_k$ can be expressed as $u_k = u_k ( \eta_k;\boldsymbol{\nu}_k, \boldsymbol{\lambda}_{k}, \boldsymbol{u}_{k-1}, \boldsymbol{a}_{k-1}, \boldsymbol{\gamma}_{k-1})$ and $a_k = a_k ( \zeta_k; \boldsymbol{\nu}_{k-1}, \boldsymbol{\lambda}_{k}, \boldsymbol{u}_{k-1}, \boldsymbol{a}_{k-1}, \boldsymbol{\gamma}_{k-1})$, respectively, where $\eta_k$ and $\zeta_k$ represent random variables that are independent of any other variables. Hence, it is possible to recursively construct $\pi^n$ with $\Prob_{\pi^n}(u_k | \boldsymbol{\nu}_k, \boldsymbol{\lambda}_{k}, \boldsymbol{u}_{k-1}, \boldsymbol{\zeta}_{k-1}, \boldsymbol{\gamma}_{k-1})$ such that it is equivalent to $\Prob_{\pi^o}(u_k | \mathcal{I}^e_k)$. This proves that $\Upsilon(\pi^n,\mu^o) = \Upsilon(\pi^o,\mu^o)$. It should be emphasized that the switching policy $\pi^n$, initially constructed associated with the control policy $\mu^o$, is now dependent solely on $\boldsymbol{\nu}_k$, $\boldsymbol{\lambda}_{k}$, $\boldsymbol{u}_{k-1}$, $\boldsymbol{\zeta}_{k-1}$, and $\boldsymbol{\gamma}_{k-1}$ at each~time~$k$.

In the second step of the proof, given the switching policy $\pi^n$, we will search for an optimal control policy $\mu^c$, and prove that it is a certainty-equivalent control policy. We first present three identities. From (3) in our paper, we have
\begin{align}\label{eq:identity1-1}
	&x_{k+1}^T S_{k+1} x_{k+1} = \Big(A_k x_k + B_k a_k + w_k \Big)^T \nonumber\\[1.5\jot]
	&\quad \qquad \qquad \qquad \quad \times S_{k+1} 	\Big(A_k x_k + B_k a_k + w_k \Big).
\end{align}
From (9) in our paper, we can write
\begin{align}\label{eq:identity1-2}
	&x_k^T S_k x_k = x_k^T \Big(Q_k + A_k^T S_{k+1} A_k \nonumber\\[1.5\jot]
	&\quad \qquad \qquad \qquad - L_k ^T \Big(B_k^T S_{k+1} B_k + R_k \Big) L_k \Big) x_k .
\end{align}
Moreover, through a simple algebraic calculation, we obtain
\begin{align}\label{eq:identity1-3}
	&x_{N+1}^T S_{N+1} x_{N+1} - x_0^T S_0 x_0 \nonumber\\[0\jot]
	&\qquad \qquad = \sum_{k=0}^{N} x_{k+1}^T S_{k+1} x_{k+1} - \sum_{k=0}^{N} x_k^T S_k x_k.
\end{align}
Incorporating the identities (\ref{eq:identity1-1}) and (\ref{eq:identity1-2}) into the identity (\ref{eq:identity1-3}), taking the expectation of both sides of (\ref{eq:identity1-3}), and using the facts that $w_k$ is independent of $x_k$ and $a_k$ and that the terms $x_0^T S_0 x_0$ and $w_k^T S_{k+1} w_k$ are independent of the switching and control policies, we find the loss function
\begin{align}\label{eq:phiprime}
	\Upsilon'(\pi,\mu) := \E \bigg[ \sum_{k=0}^{N} s_k^T \Lambda_k s_k \bigg]
\end{align}
for any $\pi \in \mathcal{P}$ and for any $\mu \in \mathcal{M}$. Note that $\Upsilon'(\pi,\mu)$ is equivalent to $\Upsilon(\pi,\mu)$ in the sense that optimizing the former over $(\pi,\mu)$ yields the same optimal solutions as optimizing the latter over $(\pi,\mu)$.

Note that, at time $k$, the term $\E[ \sum_{t=0}^{k-1} s_t^T \Lambda_t s_t ]$ in (\ref{eq:phiprime}) will not be affected by the control policy executed from time $k$ onward. Associated with $\Upsilon'(\pi^n,\mu)$ for $\pi^n$ that was obtained in the first step and for any $\mu \in \mathcal{M}$, we define the value function $V^d_k(\mathcal{I}^d_k)$ as
\begin{align}
	V^d_k(\mathcal{I}^d_k) :=& \min_{\mu \in \mathcal{M}}\E \bigg[ \sum_{t=k}^{N} s_t^T \Lambda_t s_t \Big| \mathcal{I}^d_k \bigg] \label{eq:Vd-def-proof}
\end{align}
for $k \in \mathbb{N}_{[0,N]}$ with initial condition $V^c_{N+1}(\mathcal{I}^c_{N+1}) = 0$. Building upon the prior findings in the literature (refer to, e.g., \cite{gupta2009d, khina2018t, kostina2019rate, voi, voi2}), it becomes evident that the separation principle holds, and the minimizer in (\ref{eq:Vd-def-proof}) is obtained by $a_k^\star = -L_k \hat{x}_k$. This establishes that $\Upsilon(\pi^n,\mu^c) \leq \Upsilon(\pi^n,\mu^o)$, and completes the proof of Theorem~2.

In the third step of the proof, we will show that $\Upsilon(\pi^\star,\mu^\star) \leq \Upsilon(\pi^n,\mu^c)$.  Note that, at time $k$, the term $\E[ \sum_{t=0}^{k} s_t^T \Lambda_t s_t ]$ in (\ref{eq:phiprime}) will not be affected by the switching policy executed from time $k$ onward. Associated with $\Upsilon'(\pi^n,\mu^c)$ for any $\pi^n \in \mathcal{P}$ that is innovation-based and for $\mu^c$ that was obtained in the second step, we define the value function $V^e_k(\mathcal{I}^e_k)$~as
\begin{align}
	V^e_k(\mathcal{I}^e_k) :=& \min_{\pi^n \in \mathcal{P}}\E \bigg[ \sum_{t=k+1}^{N} s_t^T \Lambda_t s_t \Big| \mathcal{I}^e_k \bigg]
\end{align}
for $k \in \mathbb{N}_{[0,N]}$ with initial condition $V^e_{N+1}(\mathcal{I}^e_{N+1}) = 0$. From the additivity of $V^e_k(\mathcal{I}^e_k)$ and by using $a_k = -L_k \hat{x}_k$, we obtain
\begin{align}\label{eq:bellman-eq}
	V^e_k(\mathcal{I}^e_k) &= \min_{\Prob(u_k|\mathcal{I}_k^e)} \E \bigg[ \hat{e}_{k+1}^T \Gamma_{k+1} \hat{e}_{k+1} \nonumber\\[1\jot]
	&\ \ +\min_{\Prob(u_{k+1}|\mathcal{I}_{k+1}^e)} \E \bigg[ \hat{e}_{k+2}^T \Gamma_{k+2} \hat{e}_{k+2} + \dots \Big|\mathcal{I}^e_{k+1} \bigg] \Big| \mathcal{I}^e_k \bigg] \nonumber\\[1\jot]
	& =\min_{\Prob(u_k|\mathcal{I}_k^e)} \E \bigg[ \hat{e}_{k+1}^T \Gamma_{k+1} \hat{e}_{k+1} + V^e_{k+1}(\mathcal{I}^e_{k+1}) \Big| \mathcal{I}^e_k \bigg] \nonumber\\[1\jot]
	& =\min_{\Prob(u_k|\mathcal{I}_k^e)} \E \bigg[ \tilde{e}_{k+1}^T \Gamma_{k+1} \tilde{e}_{k+1} + \tr(\Gamma_{k+1} P_{k+1}) \nonumber\\[1\jot]
	&\ \ + V^e_{k+1}(\mathcal{I}^e_{k+1}) \Big| \mathcal{I}^e_k \bigg]
\end{align}
for $k \in \mathbb{N}_{[0,N]}$ with initial condition $V^e_{N+1}(\mathcal{I}^e_{N+1}) = 0$, where $\Gamma_k = A_k^T S_{k+1} B_k (B_k^T S_{k+1} B_k + R_k)^{-1} B_k^T S_{k+1} A_k$, and in the third equality we used the fact that, by the tower property of conditional expectations, $\E[\hat{e}_{k+1}^T \Gamma_{k+1} \hat{e}_{k+1} | \mathcal{I}^e_k] = \E[\tilde{e}_{k+1}^T \Gamma_{k+1} \tilde{e}_{k+1} | \mathcal{I}^e_k] + \tr(\Gamma_{k+1} P_{k+1})$. We will prove by backward induction that $V^e_{k}(\mathcal{I}^e_k)$ can be written in terms of $\tilde{e}_k$, $\boldsymbol{\nu}_{k-\tau_k+1:k}$, $\tau_k$, and $\lambda_k$. The claim is satisfied for time $N+1$. We assume that the claim holds at time $k+1$, and shall prove that it also holds at time $k$.

By the hypothesis, $V^e_{k+1}(\mathcal{I}^e_{k+1})$ is function of $\tilde{e}_{k+1}$, $\boldsymbol{\nu}_{k-\tau_{k+1}+2:k+1}$, $\tau_{k+1}$, and $\lambda_{k+1}$. Note that, from (\ref{eq:mismatch-dyn}), we can write $\tilde{e}_{k+1}$ in terms of $\tilde{e}_k$, $\boldsymbol{\nu}_{k-\tau_k+1:k+1}$, $u_{k}$, $\tau_k$, and $\gamma_{k}$; from (1) in our paper, we can write $\tau_{k+1}$ in terms of $\tau_k$, $u_k$, and $\gamma_k$; and by the Markov property, we know that $\lambda_{k+1}$ depends on $\lambda_k$. Moreover, we recall that $0 \leq \tau_{k+1} \leq \tau_k +1$. Hence, there exists a function $g(.)$ such that
\begin{align}
V^e_{k+1}(\mathcal{I}^e_{k+1}) = g\Big(\tilde{e}_k, \boldsymbol{\nu}_{k-\tau_k+1:k+1}, u_k, \tau_k, \lambda_k, \gamma_k \Big).
\end{align}
This implies that $\E[V^e_{k+1}(\mathcal{I}^e_{k+1})|\mathcal{I}^e_k, u_k]$ is a function of $\tilde{e}_k$, $\boldsymbol{\nu}_{k-\tau_k+1:k}$, $u_k$, $\tau_k$, and $\lambda_k$ as $\nu_{k+1}$ and $\gamma_{k}$ are averaged out. In addition, since we can write $\tilde{e}_{k+1}$ in terms of $\tilde{e}_k$, $\boldsymbol{\nu}_{k-\tau_k+1:k+1}$, $u_{k}$, $\tau_k$, and $\gamma_{k}$, there exists a function $h(.)$ such that
\begin{align}
\tilde{e}_{k+1}^T \tilde{e}_{k+1} = h\Big(\tilde{e}_k, \boldsymbol{\nu}_{k-\tau_k+1:k+1}, u_k, \tau_k, \gamma_k \Big).
\end{align}
This implies that $\E[ \tilde{e}_{k+1}^T \tilde{e}_{k+1} | \mathcal{I}^e_k, u_k]$ is a function of $\tilde{e}_k$, $\boldsymbol{\nu}_{k-\tau_k+1:k}$, $\tau_k$, and $\lambda_k$ as $\nu_{k+1}$ and $\gamma_{k}$ are averaged out. Therefore, the claim~holds. 

Lastly, we will need to obtain the switching condition of the optimal switching policy. Note that, according to (\ref{eq:mismatch-dyn}), when $u_k = \tx$ and $\gamma_k = 1$, $\tilde{e}_{k+1}$ satisfies
\begin{align}\label{eq:mismatch-dyn-part1}
	\tilde{e}_{k+1} = K_{k+1} \nu_{k+1} .
\end{align}
When $u_k = \rtx$ and $\gamma_k = 1$, it satisfies
\begin{align}\label{eq:mismatch-dyn-part2}
	\tilde{e}_{k+1} &= \sum_{t=0}^{\tau_{k}} \bigg(\prod_{t'=1}^{t} A_{k-t'+1} \bigg) K_{k-t+1} \nu_{k-t+1} \nonumber\\[0.5\jot]
	&= \sum_{t=0}^{\tau_{k}-1} \bigg(\prod_{t'=0}^{t} A_{k-t'} \bigg) K_{k-t} \nu_{k-t} + K_{k+1} \nu_{k+1}.
\end{align}
Furthermore, when $\gamma_k = 0$, regardless of $u_k$, it satisfies
\begin{align}\label{eq:mismatch-dyn-part3}
	\tilde{e}_{k+1} = A_{k} \tilde{e}_{k} + K_{k+1} \nu_{k+1} .
\end{align}
As a result, since $\nu_{k+1}$ is a white Gaussian noise with zero mean and covariance $N_{k+1} = C_{k+1} M_{k+1} C_{k+1}^T + V_{k+1}$, from (\ref{eq:mismatch-dyn-part1}), (\ref{eq:mismatch-dyn-part2}), and (\ref{eq:mismatch-dyn-part3}), we can derive
\begin{align}
	&\E \Big[ \tilde{e}_{k+1}^T \Gamma_{k+1} \tilde{e}_{k+1} \big| \mathcal{I}^e_k, u_k = \tx, \gamma_k = 1 \Big] \nonumber\\[2\jot]
	&\quad = \tr \Big(\Gamma_{k+1} K_{k+1} N_{k+1} K_{k+1}^T \Big) \label{eq:cond-exp-mismatch-1}\\[2.5\jot]
	&\E \Big[ \tilde{e}_{k+1}^T \Gamma_{k+1} \tilde{e}_{k+1} \big| \mathcal{I}^e_k, u_k = \rtx, \gamma_k = 1 \Big] \nonumber\\[2\jot]
	&\quad = \varepsilon_k^T \Gamma_{k+1} \varepsilon_k + \tr \Big(\Gamma_{k+1} K_{k+1} N_{k+1} K_{k+1}^T \Big) \label{eq:cond-exp-mismatch-2}\\[2.5\jot]
	&\E \Big[ \tilde{e}_{k+1}^T \Gamma_{k+1} \tilde{e}_{k+1} \big| \mathcal{I}^e_k, \gamma_k = 0 \Big] \nonumber\\[2\jot]
	&\quad = \tilde{e}_k^T A_k^T \Gamma_{k+1} A_k \tilde{e}_k + \tr \Big(\Gamma_{k+1} K_{k+1} N_{k+1} K_{k+1}^T \Big) \label{eq:cond-exp-mismatch-3}
\end{align}
where in (\ref{eq:cond-exp-mismatch-2}) we have $\varepsilon_k = \sum_{t=0}^{\tau_{k}-1} (\prod_{t'=0}^{t} A_{k-t'} ) K_{k-t} \nu_{k-t}$ and used the fact that $\E[\varepsilon_k|\mathcal{I}^e_k] = \varepsilon_k$, and in (\ref{eq:cond-exp-mismatch-3}) we used the fact that $\E[\tilde{e}_k|\mathcal{I}^e_k] = \tilde{e}_k$. Employing (\ref{eq:cond-exp-mismatch-1}), (\ref{eq:cond-exp-mismatch-2}), and (\ref{eq:cond-exp-mismatch-3}), and applying the law of total expectation for the quadratic mismatch terms in (\ref{eq:bellman-eq}), we get
\begin{align}\label{eq:total-expecation-mismatch-1}
	&\E \Big[ \tilde{e}_{k+1}^T \Gamma_{k+1} \tilde{e}_{k+1} \big| \mathcal{I}^e_k, u_k = \tx \Big] \nonumber\\[3\jot]
	&\quad = \lambda_k(0) \tilde{e}_k^T A_k^T \Gamma_{k+1} A_k \tilde{e}_k \nonumber\\[3\jot]
	&\quad \quad + \lambda_k(0) \tr \Big(\Gamma_{k+1} K_{k+1} N_{k+1} K_{k+1}^T \Big) \nonumber\\[2\jot]
	&\quad \quad + \big(1-\lambda_k(0)\big) \tr \Big(\Gamma_{k+1} K_{k+1} N_{k+1} K_{k+1}^T \Big)
\end{align}
and
\begin{align}\label{eq:total-expecation-mismatch-2}
	&\E \Big[ \tilde{e}_{k+1}^T \Gamma_{k+1} \tilde{e}_{k+1} \big| \mathcal{I}^e_k, u_k = \rtx \Big] \nonumber\\[3\jot]
	&\quad = \lambda_k(\omega_k) \tilde{e}_k^T A_k^T \Gamma_{k+1} A_k \tilde{e}_k \nonumber\\[2\jot]
	&\quad \quad + \lambda_k(\omega_k) \tr \Big(\Gamma_{k+1} K_{k+1} N_{k+1} K_{k+1}^T \Big) \nonumber\\[2\jot]
	&\quad \quad + \big(1-\lambda_k(\omega_k)\big) \varepsilon_k^T \Gamma_{k+1} \varepsilon_k \nonumber\\[2\jot]
	&\quad \quad + \big(1-\lambda_k(\omega_k)\big) \tr \Big(\Gamma_{k+1} K_{k+1} N_{k+1} K_{k+1}^T \Big).
\end{align}

Besides, applying the law of total expectation for the cost-to-go terms in (\ref{eq:bellman-eq}), we get
\begin{align}\label{eq:total-expecation-costtogo-1}
	&\E \Big[ V^e_{k+1}(\mathcal{I}^e_{k+1}) \big| \mathcal{I}^e_k, u_k = \tx \Big] \nonumber\\[2\jot]
	& = \lambda_k(0) \E \Big [ V^e_{k+1}(\mathcal{I}^e_{k+1}) \big| \mathcal{I}^e_k, u_k = \tx, \gamma_k = 0 \Big] \nonumber\\[2\jot]
	& + \big(1 \!- \! \lambda_k(0)\big) \E \Big [ V^e_{k+1}(\mathcal{I}^e_{k+1}) \big| \mathcal{I}^e_k, u_k = \tx, \gamma_k = 1 \Big]
\end{align}
and
\begin{align}\label{eq:total-expecation-costtogo-2}
	&\E \Big[ V^e_{k+1}(\mathcal{I}^e_{k+1}) \big| \mathcal{I}^e_k, u_k = \rtx \Big] \nonumber\\[2\jot]
	& = \lambda_k(\omega_k) \E \Big [ V^e_{k+1}(\mathcal{I}^e_{k+1}) \big| \mathcal{I}^e_k, u_k = \rtx, \gamma_k = 0 \Big] \nonumber\\[2\jot]
	& + \big(1 \!- \! \lambda_k(\omega_k)\big) \E \Big [ V^e_{k+1}(\mathcal{I}^e_{k+1}) \big| \mathcal{I}^e_k, u_k = \rtx, \gamma_k = 1 \Big].
\end{align}
Observe that since no measurement can be detected correctly at time $k+1$ when $\gamma_k = 0$, in (\ref{eq:total-expecation-costtogo-1}) and (\ref{eq:total-expecation-costtogo-2}), we can use the equalities
\begin{align*}
&\E \Big[ V^e_{k+1}(\mathcal{I}^e_{k+1}) \big| \mathcal{I}^e_k, u_k = \tx, \gamma_k = 0 \Big]\\[2\jot]
&\quad = \E \Big[V^e_{k+1}(\mathcal{I}^e_{k+1}) \big| \mathcal{I}^e_k, u_k = \rtx, \gamma_k = 0 \Big]\\[2\jot]
&\quad = \E \Big [V^e_{k+1}(\mathcal{I}^e_{k+1}) \big| \mathcal{I}^e_k, \gamma_k = 0 \Big].
\end{align*}

Inserting (\ref{eq:total-expecation-mismatch-1}), (\ref{eq:total-expecation-mismatch-2}), (\ref{eq:total-expecation-costtogo-1}), and (\ref{eq:total-expecation-costtogo-2}) in (\ref{eq:bellman-eq}), we deduce that $u_k^\star = \tx$ if
\begin{align*}
&\big(\lambda_k(\omega_k) - \lambda_k(0)\big) \tilde{e}_k^T A_k^T \Gamma_{k+1} A_k \tilde{e}_k \nonumber\\[3.25\jot]
& + \big(1- \lambda_k(\omega_k)\big) \varepsilon_k^T \Gamma_{k+1} \varepsilon_k \nonumber\\[2.75\jot]
& + \big(\lambda_k(\omega_k) - \lambda_k(0)\big) \E \Big[ V^e_{k+1}(\mathcal{I}^e_{k+1}) \big| \mathcal{I}^e_k, \gamma_k = 0 \Big] \nonumber\\[2\jot]
& - \big(1 - \lambda_k(0)\big) \E \Big[ V^e_{k+1}(\mathcal{I}^e_{k+1}) \big| \mathcal{I}^e_k, u_k = \tx, \gamma_k = 1 \Big] \nonumber\\[2\jot]
& + \big(1 - \lambda_k(\omega_k)\big) \E \Big[ V^e_{k+1}(\mathcal{I}^e_{k+1}) \big| \mathcal{I}^e_k, u_k = \rtx, \gamma_k = 1 \Big] \geq 0 .
\end{align*}
We also know that $u_k^\star = \tx$ if $\omega_k > \omega_{\max}$ or $\tau_k = 0$. Note that the condition $\omega_k > \omega_{\max}$ comes from the HARQ scheme, and the condition $\tau_k = 0$ from the fact that the previous transmission was successful. In both cases, no retransmission is taken place. This completes the proof of Theorem~1.
\end{IEEEproof}

\bibliography{mybib}
\bibliographystyle{ieeetr}

\end{document}